\documentclass[twocolumn,aps,prb,superscriptaddress,amsmath,amssymb,superscriptaddress]{revtex4-1}
\newcommand{\pagenumbaa}{1}
\usepackage{graphicx}
\usepackage{color}
\usepackage{braket}

\usepackage{verbatim}
\usepackage{amsmath}
\usepackage{mathtools}

\usepackage{float}
\usepackage{hyperref}
\usepackage{lipsum}
\usepackage{upgreek}

\begin{document}
\title{Low-Noise GaAs Quantum Dots for Quantum Photonics}
\author{Liang Zhai}
\email{liang.zhai@unibas.ch}  
\affiliation{Department of Physics, University of Basel, Klingelbergstrasse 82, CH-4056 Basel, Switzerland}
\author{Matthias C. L\"obl}
\affiliation{Department of Physics, University of Basel, Klingelbergstrasse 82, CH-4056 Basel, Switzerland}
\author{Giang N. Nguyen}
\affiliation{Department of Physics, University of Basel, Klingelbergstrasse 82, CH-4056 Basel, Switzerland}
\affiliation{Lehrstuhl f\"ur Angewandte Festk\"orperphysik, Ruhr-Universit\"at Bochum, DE-44780 Bochum, Germany}
\author{Julian Ritzmann}
\affiliation{Lehrstuhl f\"ur Angewandte Festk\"orperphysik, Ruhr-Universit\"at Bochum, DE-44780 Bochum, Germany}
\author{Alisa Javadi}
\affiliation{Department of Physics, University of Basel, Klingelbergstrasse 82, CH-4056 Basel, Switzerland}
\author{Clemens Spinnler}
\affiliation{Department of Physics, University of Basel, Klingelbergstrasse 82, CH-4056 Basel, Switzerland}
\author{Andreas D. Wieck}
\affiliation{Lehrstuhl f\"ur Angewandte Festk\"orperphysik, Ruhr-Universit\"at Bochum, DE-44780 Bochum, Germany}
\author{Arne Ludwig}
\affiliation{Lehrstuhl f\"ur Angewandte Festk\"orperphysik, Ruhr-Universit\"at Bochum, DE-44780 Bochum, Germany}
\author{Richard J. Warburton}
\affiliation{Department of Physics, University of Basel, Klingelbergstrasse 82, CH-4056 Basel, Switzerland}

\begin{abstract}
Quantum dots are both excellent single-photon sources and hosts for single spins. This combination enables the deterministic generation of Raman-photons -- bandwidth-matched to an atomic quantum-memory -- and the generation of photon cluster states, a resource in quantum communication and measurement-based quantum computing. GaAs quantum dots in AlGaAs can be matched in frequency to a rubidium-based photon memory, and have potentially improved electron spin coherence compared to the widely used InGaAs quantum dots. However, their charge stability and optical linewidths are typically much worse than for their InGaAs counterparts. Here, we embed GaAs quantum dots into an $n$-$i$-$p$-diode specially designed for low-temperature operation. We demonstrate ultra-low noise behaviour: charge control via Coulomb blockade, close-to lifetime-limited linewidths, and no blinking. We observe high-fidelity optical electron-spin initialisation and long electron-spin lifetimes for these quantum dots. Our work establishes a materials platform for low-noise quantum photonics close to the red part of the spectrum.
\end{abstract}

\maketitle

\setcounter{page}{\pagenumbaa}
\thispagestyle{plain}

Quantum dots (QDs) in III-V semiconductors form excellent sources of indistinguishable single-photons. These emitters have a combination of metrics (brightness, purity, coherence, repetition rate) which no other source can match \cite{Somaschi2016,Ding2016,Liu2018,Liu2019}. These excellent photonic properties can be extended by trapping a single electron to the QD, enabling spin-photon entanglement \cite{Gao2012} and high-rate remote spin-spin entanglement creation \cite{Stockill2017}. Underpinning these developments are, first, a self-assembly process to create nano-scale QDs; and second, a smart heterostructure design along with high-quality material. The established platform consists of InGaAs QDs embedded in GaAs. However, the InGaAs QDs emit at wavelengths between 900 nm and 1200 nm, a spectral regime lying inconveniently between the telecom wavelengths (1300 nm and 1550 nm) and the wavelength where silicon detectors have a high efficiency\cite{Sangouard2011} (600 nm - 800 nm). It is important in the development of QD quantum photonics to extend the wavelength range towards both, shorter and longer wavelengths.

GaAs QDs in an AlGaAs matrix can be self-assembled by local droplet etching \cite{Huo2013,Gurioli2019} and have a spectrally narrow ensemble \cite{Heyn2009,Lobl2019}. They emit at wavelengths between 700--800 nm. This is an important band: it coincides with the peak quantum efficiency of silicon detectors; it contains the rubidium D$_1$ and D$_2$ wavelengths (795 nm and 780 nm, respectively) offering a powerful route to combining QD photons with a rubidium-based quantum memory \cite{Wolters2017}. Furthermore, GaAs QDs have typically more symmetric shapes, facilitating the creation of polarisation-entangled photon pairs from the biexciton cascade \cite{Liu2019,Basset2018}.

GaAs QDs have also very low levels of strain \cite{Urbaszek2013,Plumhof2010,Gurioli2019,Neul2015,Phy9}. In contrast, the high level of strain in InGaAs QDs complicates the interaction of an electron spin with the nuclear spins on account of the atomic site-specific quadrupolar interaction \cite{Urbaszek2013,Gangloff2019}. For electrostatically defined GaAs QDs, the spin-dephasing time, $T_{2}^{*}$, has been prolonged to the micro-second regime by narrowing the nuclear spin distribution together with real-time Hamiltonian estimation\cite{Shulman2014}. Applied to a droplet GaAs QD, such techniques could prolong the spin dephasing time to values several orders of magnitude above the radiative lifetime. In this case, in combination with optical cavities \cite{Najer2019}, droplet GaAs QDs can potentially serve as fast, high-fidelity sources of spin-photon pairs and cluster states \cite{Schwartz2016}.

\begin{figure*}[t]
\includegraphics[width=1.8\columnwidth]{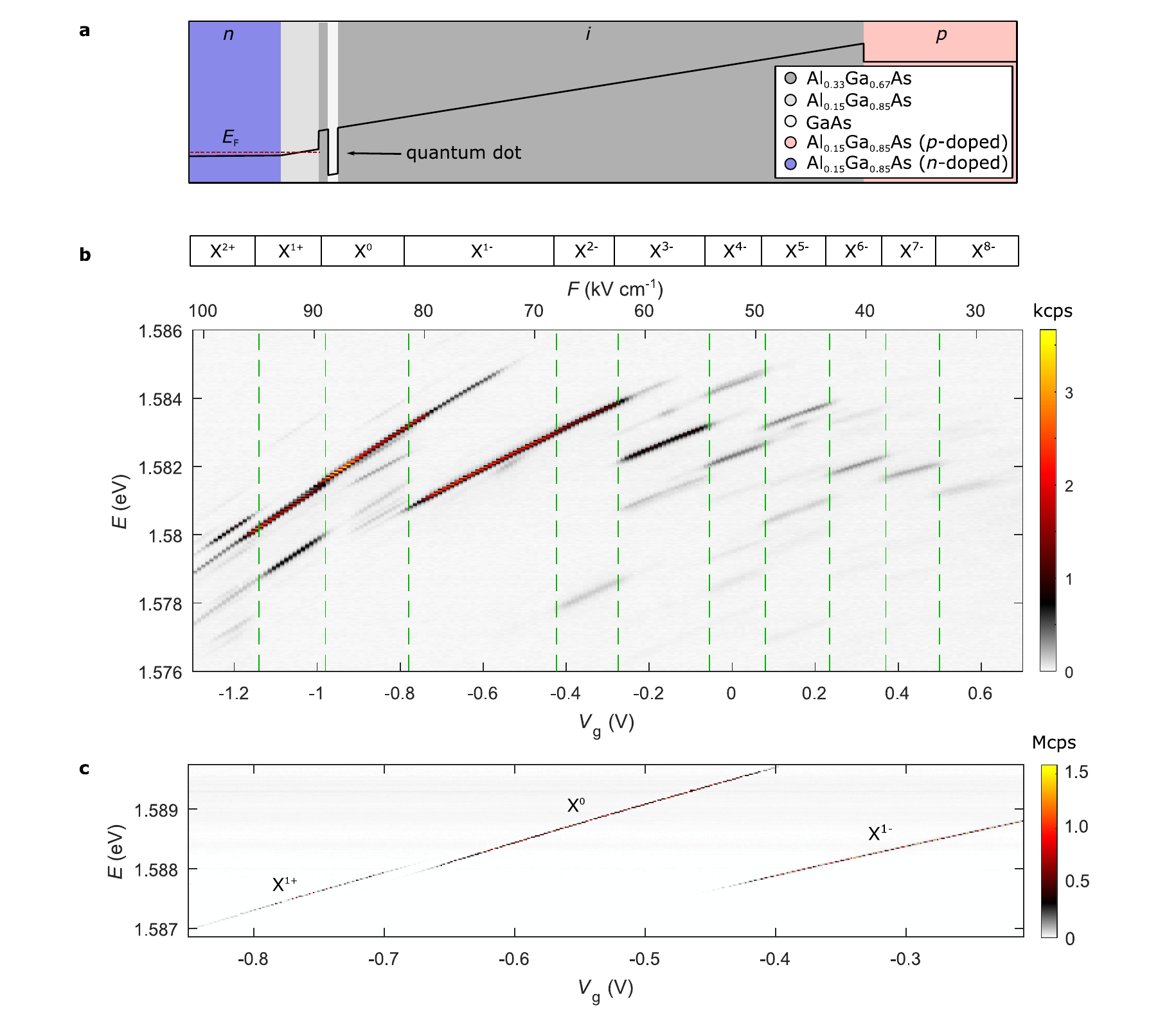}
\caption{\label{fig:chargeTune}\textbf{Tuning the charge state of single GaAs quantum dots. (a)} Schematic band structure (conduction band) of the diode hosting charge-tunable GaAs quantum dots. \textbf{(b)} The photoluminescence emitted by an exemplary single quantum dot as a function of the gate voltage, $V_{\text{g}}$. (Positive gate voltage indicates a forward bias.) The corresponding electric field, $F$, is plotted as an additional $x$-axis on top. The photoluminescence is resolved in energy by a spectrometer and measured on a CCD-camera. The emission spectrum shows several plateaus corresponding to different charge states of the quantum dot. We observe narrow photoluminescence-linewidths on highly charged excitons where up to eight additional electrons occupy the quantum dot. \textbf{(c)} Resonance fluorescence from X$^{1+}$, X$^{0}$, and X$^{1-}$ charge plateaus measured on another quantum dot (QD1). X$^{1+}$,  X$^{0}$, and X$^{1-}$ represent the positive trion, the neutral exciton, and the negative trion, respectively. The measurement is performed by sweeping the gate voltage for different laser frequencies. The resonance fluorescence intensity is measured with a superconducting nanowire single-photon detector. This measurement is performed by resonant continuous-wave excitation below saturation. In saturation, the maximum count rate is $6.5\ $MHz (see Supplementary Figure\ 1 for the power saturation curve).}
\end{figure*}

The development of GaAs QDs for quantum photonics lags far behind the InGaAs QDs. Recurrent problems are blinking \cite{Jahn2015,Beguin2018} (telegraph noise in the emission) and optical linewidths well above the transform limit \cite{Kumar2011,Neul2015,Beguin2018,Basset2018,Scholl2019}. Both of these problems are caused by charge noise. On short time-scales, the charge environment is static such that successively emitted photons exhibit a high degree of coherence \cite{Scholl2019,Liu2019}. On longer time-scales, however, the charge noise introduces via blinking an unacceptable stochastic character to the photon stream. An additional weak non-resonant laser provides control over the noise to a certain extend, though it does not remove the blinking completely\cite{Jahn2015}.

\begin{figure*}[t]
\includegraphics[width=2.0\columnwidth]{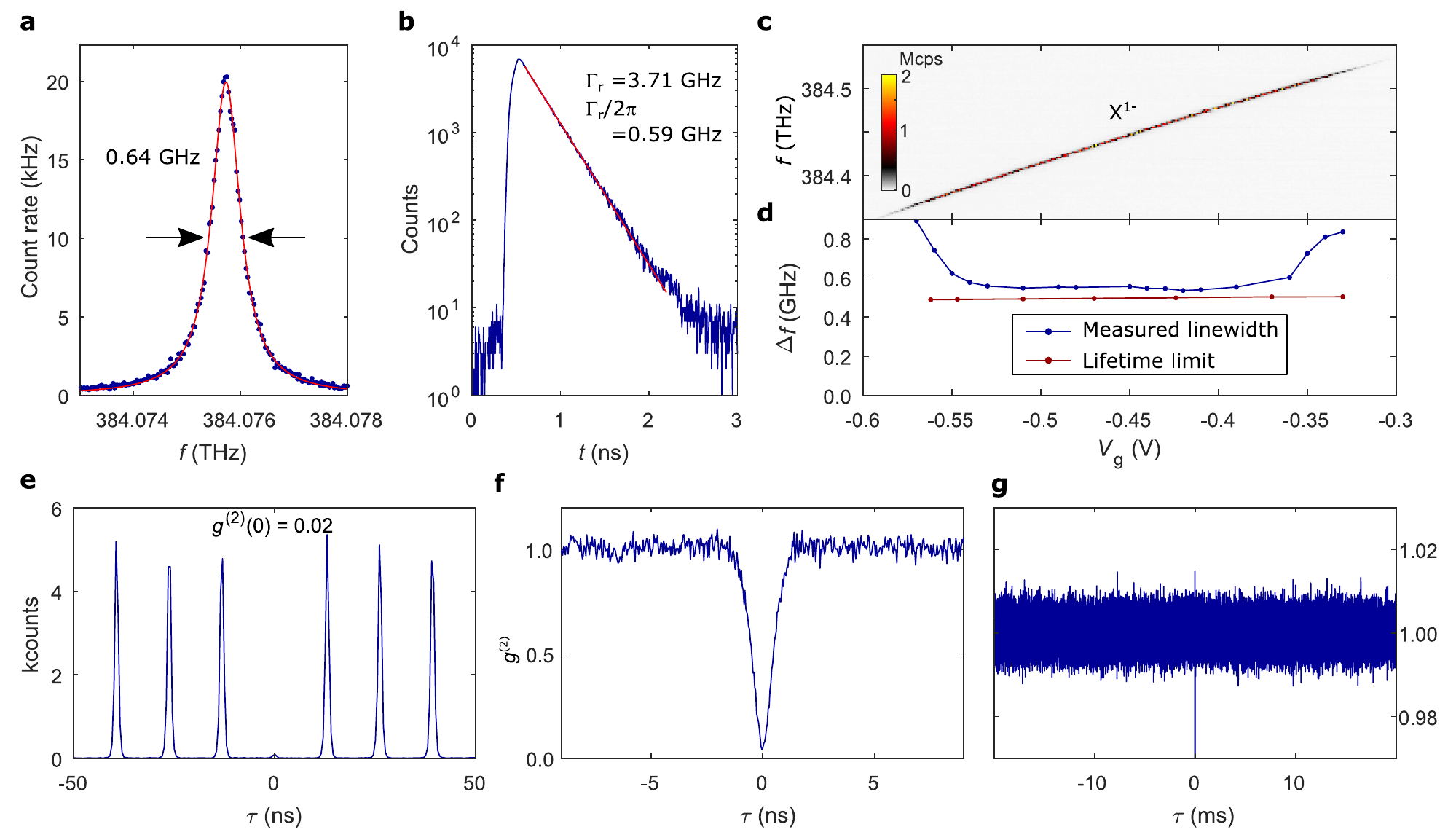}
\caption{\label{fig:Linewidth}{\bf Time-resolved lifetime and photon-correlation measurements. (a)} Resonance fluorescence linewidth measured on the singly-charged exciton, X$^{1-}$ (QD1). The measurement is performed by sweeping a narrow-bandwidth laser over the X$^{1-}$ resonance. The overall time for the shown scan is $\sim8$ min. A Lorentzian function (red line) fits perfectly to the data (blue dots), showing an optical linewidth of $0.64 \pm 0.01\ \text{GHz}$. \textbf{(b)} Lifetime measurement on X$^{1-}$ under pulsed resonant excitation. The gate voltage is the same as in (a). The measured decay rate ($\Gamma_{\text{r}}=3.71 \pm 0.04\ \text{GHz}$, corresponding to a lifetime of $1/\Gamma_{\text{r}}=270 \pm 3\ \text{ps}$) implies a lifetime-limited linewidth of $\Gamma_{\text{r}}/2\uppi = 0.59 \pm 0.01\ \text{GHz}$ (Exponential fit). \textbf{(c)} Resonance fluorescence of X$^{1-}$ (QD2) as a function of the gate voltage. \textbf{(d)} Resonance fluorescence linewidth along with the lifetime-limit (obtained from separate lifetime measurements at the corresponding gate voltages). Similar to QD1, the linewidth of QD2 stays very close to the lifetime limit in the plateau centre. {\bf (e)} Auto-correlation ($g^{(2)}$) measured under resonant $\uppi$-pulse excitation. \textbf{(f)} Auto-correlation of the resonance fluorescence measured under weak continuous-wave excitation shown on a short time-scale. The $g^{(2)}$-measurement is normalised\cite{Lobl2020} by dividing the number of coincidences by its expectation value $T\cdot t_{\text{bin}}\cdot x_1\cdot x_2$, where $T$ is the overall integration time, $t_{\text{bin}}$ is the binning time, and $x_1$, $x_2$ are the count-rates on the two single-photon detectors. \textbf{(g)} The same auto-correlation measurement as in (f) but evaluated on a much longer time-scale (milliseconds). The perfectly flat $g^{(2)}$ reveals the absence of blinking.}
\end{figure*}

\begin{figure*}[t]
\includegraphics[width=2.0\columnwidth]{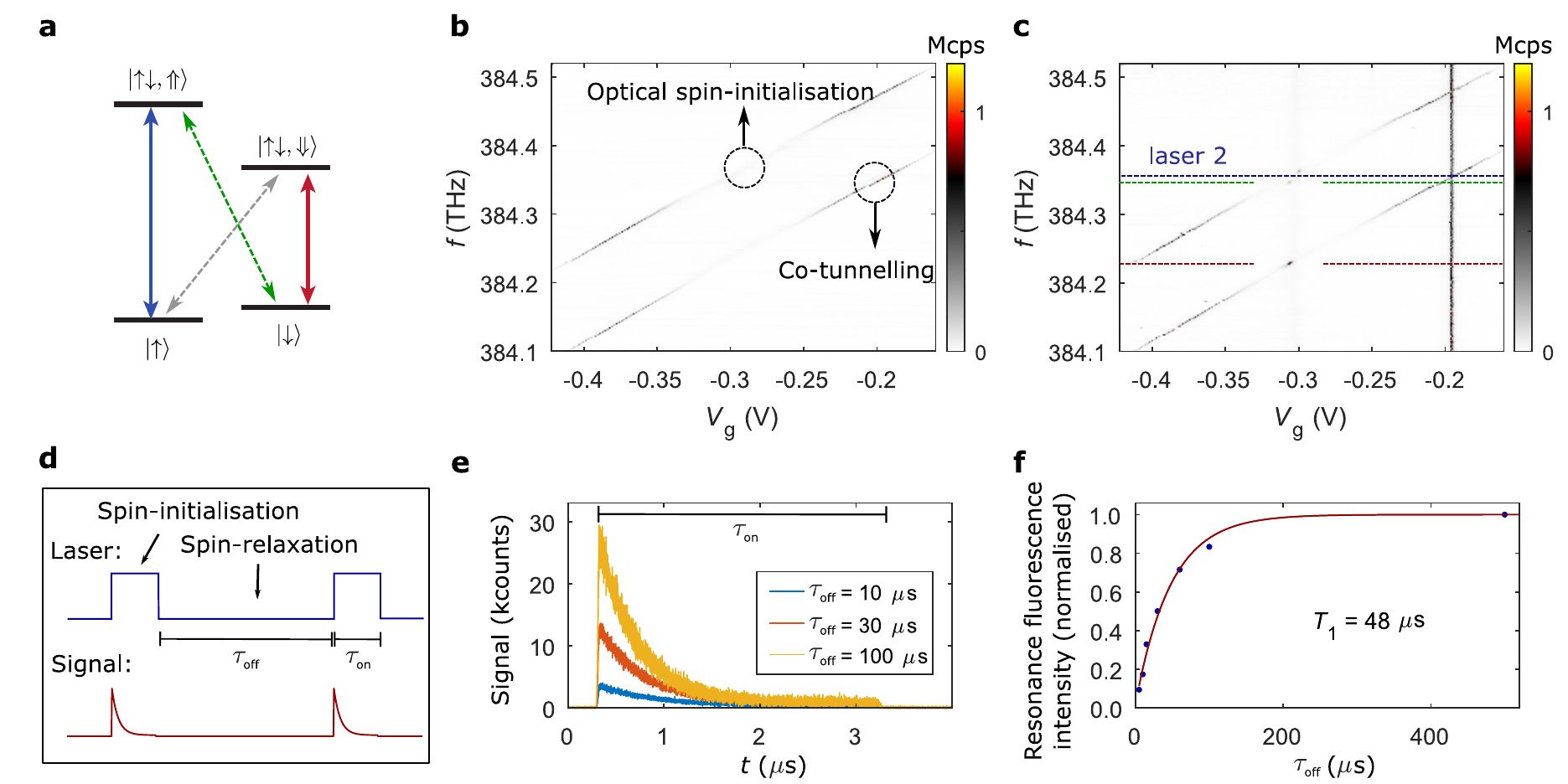}
\caption{\label{fig:spinPump}{\bf Initialisation of a single electron spin. (a)} Level scheme of the negative trion X$^{1-}$ in a magnetic field (Faraday geometry). \textbf{(b)} Optical spin-initialisation via optical pumping on X$^{1-}$. The measurement is carried out at $B=6.6$ T. In the plateau centre, the resonance fluorescence disappears due to successful spin-initialisation; at the plateau edges it remains bright due to rapid spin-randomisation via co-tunneling \cite{Smith2005}. \textbf{(c)} Optical spin-initialisation and re-pumping with a second laser at a fixed frequency (laser 2). Recoveries of the signal are found in the plateau centre. \textbf{(d)} Schematic of the time-resolved spin-pumping measurement. \textbf{(e)} Resonance fluorescence intensity as a function of time. The signal drops due to optical spin-initialisation after turning the driving laser on. The overall intensity is larger when the time-delay $\tau_{\text{off}}$ between the laser pulses is larger. In this case, the electron spin has more time to relax back from the off-resonant state. \textbf{(f)} Resonance fluorescence intensity as a function of the waiting time between the spin-pumping laser pulses. The magenta line is an exponential fit to the data (blue dots). From this measurement we extract an electron-spin lifetime of $T_1\sim48\pm5\ \mu$s.}
\end{figure*}

For InGaAs QDs, embedding the QDs in an $n$-$i$-$p$ diode has profound advantages: the charge state is locked by Coulomb blockade \cite{Warburton2000,Lobl2017,Grim2019}; the charge noise is reduced significantly\cite{Kuhlmann2014}; and the exact transition frequency can be tuned in-situ via a gate voltage \cite{Liu2018,Patel2010}. Such a structure is missing for GaAs QDs \cite{Jahn2015,Beguin2018,Neul2015,Kumar2011,Basset2018,Scholl2019} -- in previous attempts, charge-stability was not demonstrated \cite{Langer2014,Bouet2014}. A materials issue must be addressed: the barrier material AlGaAs must be doped, yet silicon-doped AlGaAs contains DX-centres \cite{Mooney1990,Munoz1993} which both reduce the electron concentration, causing the material to freeze out at low temperatures, and lead to complicated behaviour under illumination. Here, we resolve this issue -- all doped AlGaAs layers have a low Al-concentration. In this case, the DX level lies above the conduction band minimum and thus is unoccupied at cryogenic temperatures\cite{Mooney1990}. The QDs are grown in a region with higher Al-concentration, which is well-established for the growth of these QDs \cite{Huo2013}. On GaAs QDs in this device we demonstrate charge-control via Coulomb blockade, optical linewidths just marginally above the transform limit, blinking-free single-photon emission, electron spin initialisation, and a spin-relaxation time as large as $\sim50\ \mu$s.

\section*{Results}
\subsection*{Sample design and characterisation}
The sample is grown on a GaAs-substrate with (001)-orientation. Below the active region of the sample, a distributed Bragg reflector is grown to enhance the collection efficiency of the photons emitted by the QDs. The QDs are embedded in an $n$-$i$-$p$-diode structure where the QDs are tunnel-coupled to the $n$-type layer. The $n$-type back gate consists of silicon-doped Al$_{0.15}$Ga$_{0.85}$As. The low Al-concentration in this layer is crucial to avoid the occupation of DX-centres in $n$-type AlGaAs \cite{Mooney1990,Munoz1993}. A tunnel barrier consisting of 20 nm Al$_{0.15}$Ga$_{0.85}$As followed by 10 nm Al$_{0.33}$Ga$_{0.67}$As separates the QDs from the $n$-type back gate. The QDs are grown in the Al$_{0.33}$Ga$_{0.67}$As-layer by using local droplet-etching \cite{Huo2013}. The QD-density is $n_{\text{QD}}=0.37\pm0.01\ \mu$m$^{-2}$. Above the QDs, there is 274 nm of Al$_{0.33}$Ga$_{0.67}$As followed by a $p$-type top gate. The top gate is composed of carbon-doped Al$_{0.15}$Ga$_{0.85}$As, where reduced Al-concentration is used as well. A schematic bandstructure of the diode is shown in Fig.\ \ref{fig:chargeTune}(a); all Al-concentrations in this design are small enough that processing into micropillars\cite{Bockler2008} and nanostructures will not be hindered by oxidation\cite{Kirsanske2017}. In Table\ \ref{table_structure}, details of the full heterostructure are given.

We characterise our device by measuring the photoluminescence from a single QD as a function of the gate voltage, $V_{\text{g}}$, applied to the diode (Fig.\ \ref{fig:chargeTune}(b)). As a function of $V_{\text{g}}$, the emission lines show a pronounced Stark-shift. At specific gate voltages, discrete jumps in the emission spectrum take place: one emission line abruptly becomes weaker and another line appears. This effect is the characteristic signature of charge-control of a QD via Coulomb blockade \cite{Warburton2000}: the net-charge of the QD increases one by one and the emission energy is shifted due to the additional Coulomb interaction with the new carrier. 

We fit the relation $E=E_0+\alpha F+\beta F^2$ to the dependence of the emission energy, $E$, on electric field, $F$ (Supplementary Figure\ 2). The energy jumps between different charge plateaus are removed for the fit. We find $\alpha/\text{e}=0.21\ \text{nm}$, the permanent dipole moment in the growth direction, and $\beta=-1.35\cdot10^{-6}\ \text{eV}\ (\text{kV}/\text{cm})^{-2}$, the polarisability of the QD \cite{Fry2000}. Extrapolating the fit shows that the Stark shift is zero at a non-zero electric field ($F=7.8\ \text{kV}\ \text{cm}^{-1}$). The non-zero value of $\alpha$ represents a small displacement between the ``centre-of-mass" of the electron and the hole wavefunctions. The hole wavefunction is slightly closer to the back gate than the electron wavefunction.
\subsection*{Resonance fluorescence from GaAs QDs}
We identify the neutral exciton, X$^{0}$, from its characteristic fine-structure splitting as well as a quantum-beat in time-resolved resonance fluorescence (Supplementary Figure\ 3). For our device, the fine-structure splittings are distributed over a range of $1-3\ $GHz (see Supplementary Figure\ 4(c)). The fine-structure splittings are comparable to literature values on (001)-oriented samples \cite{Huo2013,Liu2019}. Smaller fine-structure splittings can be obtained by using (111)-oriented samples\cite{Basset2018} and strain-tuning\cite{Huber2018}. We identify the other charge-states by counting the number of jumps in the emission spectrum as the gate-voltage increases/decreases. We measure emission from highly charged excitons ranging from the two-times positively charged exciton, X$^{2+}$, to the eight-times negatively charged exciton, X$^{8-}$. Such a wide range of charge tuning was not previously achieved with any QDs emitting in the close-to-visible wavelengths. Our GaAs QDs give a large range of charge tuning due to their relatively large size \cite{Huo2013} in comparison to the widely used InGaAs QDs \cite{Madsen2013}.

We turn to resonant excitation. This excitation scheme is key for creating low-noise photons and represents a true test of the fidelity of the device as, unlike photoluminescence, continuum states are not deliberately occupied. By sweeping both the gate voltage and excitation laser frequency, we map out three charge plateaus of a single quantum dot (QD1) -- X$^{1+}$, X$^{0}$, and X$^{1-}$ (see Supplementary Figure\ 5 for photoluminescence of QD1). As is visible in Fig.\ \ref{fig:chargeTune}(c), the exact transition energy of all three charge states can be tuned via $V_{\text{g}}$ across a range of above $1\ \text{meV}$. At a fixed gate voltage, we determine a resonance fluorescence linewidth of X$^{1-}$ to be $0.64 \pm 0.01$ GHz (full width at half maximum) on scanning a narrow-bandwidth laser over the trion resonance (see Fig.\ \ref{fig:Linewidth}(a)). (resonance fluorescence laser scans on X$^{1+}$ and X$^{0}$ are shown in Supplementary Figure\ 3). This measurement takes several minutes: the linewidth probes the sum of all noise sources over an enormous frequency bandwidth\cite{Kuhlmann2013}. The measured linewidth is very close to the lifetime-limit of $\Gamma_{\text{r}}/2\uppi=0.59 \pm 0.01 \ \text{GHz}$. (It is assumed here the decay is radiative. The radiative decay rate $\Gamma_{\text{r}}$ is determined by recording a decay curve following pulsed resonant excitation, Fig.\ \ref{fig:Linewidth}(b)). This result shows that there is extremely little linewidth broadening due to noise in our device. These excellent results are not limited to one individual QD. Shown in Fig.\ \ref{fig:Linewidth}(d) is a linewidth measurement on a second QD (QD2). In the central part of the X$^{1-}$ charge-plateau (from $V_{\text{g}}=-0.5\ $V to $V_{\text{g}}=-0.4\ $V  in Fig.\ \ref{fig:Linewidth}(c)), we also measure a close-to lifetime-limited linewidth. On average, the ratio between the measured linewidth and the lifetime limit is $1.08$ for QD2. At the edges of the charge-plateau, the linewidth increases -- a well-know effect due to a co-tunneling interaction with the Fermi-reservoir \cite{Smith2005}. Comparably good properties are found for in total seven out of ten randomly chosen QDs with X$^{1-}$ below $785\ $nm (see Supplementary Figure\ 4(a,b)).

A remarkable feature is that the close-to-transform limited linewidths are observed despite the large dc Stark shifts of these QDs. Within the X$^{1-}$ plateau of QD1 (Fig.\ \ref{fig:chargeTune}(c)), the dc Stark shift is 
$0.0347\ \text{GHz}$ per $\text{V}\ \text{cm}^{-1}$, about a factor of four larger than the typical dc Stark shifts of InGaAs QDs \cite{Kuhlmann2013}. The sensitivity of the transition frequency to the electric field renders the QD linewidth susceptible to charge noise. The close-to-transform limited linewidths reflect therefore an extremely low level of charge noise in the device. Assuming that the slight increase in broadening with respect to the transform limit arises solely from charge noise, the linewidth measurement places an upper bound of $\sim3.0\ \text{V}\ \text{cm}^{-1}$ for the root-mean-square (rms) electric field noise at the location of QD1. This upper bound is comparable to the best gated InGaAs QD devices. \cite{Kuhlmann2013,Kuhlmann2014,Matthiesen2012,Lu2010,Najer2019}.

For applications as single-photon source, it is crucial to demonstrate that the photons are emitted one by one, i.e.\ photon anti-bunching. Therefore, we continue our analysis by performing an intensity auto-correlation of the resonance fluorescence. This $g^{(2)}$-measurement is shown in Fig.\ \ref{fig:Linewidth}(e) and Supplementary Figure\ 6(c,d) for resonant $\uppi$-pulse excitation with 76 MHz repetition rate. We observe a strong anti-bunching at zero time delay ($g^{(2)}(0)=0.019\pm0.008$), corresponding to a single-photon purity of $1-g^{(2)}(0)\sim98\%$. The corresponding measurement under weak continuous-wave excitation is shown in Fig.\ \ref{fig:Linewidth}(f). ($g^{(2)}$-measurements versus excitation power as well as laser detuning are mapped out in Supplementary Figure\ 7, where clear Rabi oscillations are shown. In both cases, we find excellent agreement between the measured $g^{(2)}$ and a calculation based on a two-level model.) Also here, we observe a strong anti-bunching proving the single-photon nature of the emission. 

Previous resonance fluorescence on GaAs QDs has suffered from blinking, i.e. telegraph noise in the emission \cite{Jahn2015}. This is a deleterious consequence of charge noise: either the QD charges abruptly or the charge state of a nearby trap changes, detuning the QD from the excitation laser in both cases. Blinking gives rise to a characteristic bunching ($g^{(2)}>1$) in the auto-correction even for driving powers well below saturation\cite{Jahn2015}. We investigate this point here. Even out to long (millisecond) time-scales, the $g^{(2)}$-measurement is absolutely flat and close to one (see Fig.\ \ref{fig:Linewidth}(g)). (We note that our analysis includes a mathematically justified normalisation of the $g^{(2)}$-measurement \cite{Lobl2020}.) This result demonstrates that blinking is absent. This is a consequence both of the diode-structure, in particular Coulomb blockade which locks the QD charge, and the low charge noise in the material surrounding the QD. 

We subsequently carried out $g^{(2)}$-measurements with either a small magnetic field along the growth direction or a laser slightly detuned from the QD resonance. In the former case the sensitivity to spin noise is enhanced, while in the latter case the sensitivity to charge noise is enhanced\cite{Kuhlmann2013}. In Supplementary Figure\ 8, we compare the $g^{(2)}$-measurements on millisecond time-scales. For the measurement with an additional magnetic field (Supplementary Figure\ 8(c,d)), the $g^{(2)}$ remains flat and stays close to one. In contrast, we observe a small blinking when the laser is detuned (Supplementary Figure\ 8(e,f)). We infer from these results that in our device charge noise is most likely to be responsible for the residual linewidth broadening.

\subsection*{High-fidelity spin initialisation}
The diode structure allows us to load a QD with a single electron. The spin of the electron is a valuable quantum resource. To probe the electron-spin dynamics, we probe the X$^{1-}$ resonance fluorescence in a magnetic field (Faraday-geometry). In this configuration, the ground state is split by the electron Zeeman energy, and the excited state is split by the hole Zeeman energy (see Fig.\ \ref{fig:spinPump}(a)). As the diagonal transitions in this level-scheme are close-to forbidden, the X$^{1-}$-charge-plateau splits into two lines which are separated by the sum of electron and hole Zeeman energies (see Fig.\ \ref{fig:spinPump}(b)). We find that the X$^{1-}$ charge-plateau becomes optically dim in its centre. This is the characteristic feature of spin-initialisation via optical pumping \cite{Atature2006,Lu2010,Lobl2017,Javadi2017}. On driving e.g.\ the $\ket{\uparrow} - \ket{\uparrow\downarrow\Uparrow}$ transition, the trion will most likely decay back to the $\ket{\uparrow}$-state via the dipole-allowed vertical transition. However, due to the heavy-hole light-hole mixing or a weak in-plane nuclear field, it can also decay to the $\ket{\downarrow}$-state through the ``forbidden" transtion with a small probability. When the QD is in the $\ket{\downarrow}$-state, the driving laser is off-resonance on account of the electron Zeeman energy. Therefore, the centre of the X$^{1-}$-charge-plateau becomes dark and the initialisation of the electron spin in the $\ket{\downarrow}$-state is heralded by the disappearing resonance fluorescence. At the plateau-edges, resonance fluorescence reappears due to fast spin-randomisation via co-tunneling \cite{Smith2005}. By comparing the remaining intensity in the charge-plateau centre to the plateau edges\cite{Lobl2017}, we estimate the spin initialisation fidelity to be $F = 98.3\pm0.3\ \%$. To confirm that the signal disappears in the plateau-centre on account of optical spin initialisation and not some other process, we perform a measurement with a second laser at a fixed frequency. When the fixed laser is resonant with $\ket{\uparrow}-\ket{\uparrow\downarrow\Uparrow}$ transition, we observe a recovery of the signal (Fig.\ \ref{fig:spinPump}(c)) on either driving the weak diagonal transition $\ket{\downarrow}-\ket{\uparrow\downarrow\Uparrow}$ or the strong vertical transitions $\ket{\downarrow}-\ket{\uparrow\downarrow\Downarrow}$ with the scan laser. While the fixed laser is tuned to $\ket{\downarrow}-\ket{\uparrow\downarrow\Uparrow}$ transition (at a different $V_{\text{g}}$), another recovery spot is seen as the scan laser drives the vertical transition $\ket{\uparrow}-\ket{\uparrow\downarrow\Uparrow}$. This confirms the optical spin-initialisation mechanism \cite{Atature2006,Lobl2017}. From the energy splitting at the plateau edges, we determine the electron and hole g-factors\cite{Phy9}, $g_{\text{e}} = -0.076\pm0.001$ and $g_{\text{h}} = 1.309\pm0.001$. For the positively charged trion (X$^{1+}$), we also observe high-fidelity optical spin-initialisation (Supplementary Figure\ 9) and narrow linewidths ($0.62$ GHz, see Supplementary Figure\ 3), in this case of a hole spin.

How long-lived is the prepared spin state? To answer this question, we measure the time-dependence of the X$^{1-}$ spin initialisation \cite{Lu2010,Javadi2017}. The scheme is illustrated in Fig.\ \ref{fig:spinPump}(d). First, we drive the $\ket{\uparrow}-\ket{\uparrow\downarrow\Uparrow}$ transition for $\tau_{\text{on}}=3\ \mu$s. During this laser pulse, the signal decreases due to optical spin-initialisation (Fig.\ \ref{fig:spinPump}(e)). Subsequently, we turn the laser off for a time $\tau_{\text{off}}$, and then turn the laser back on again. During the off-time the electron spin randomises. Fig.\ \ref{fig:spinPump}(e) shows that the resonance fluorescence signal is stronger when the waiting time $\tau_{\text{off}}$ is longer. The reason for this effect is that with increasing $\tau_{\text{off}}$ the spin has more time to randomise. For a short value of $\tau_{\text{off}}$, in contrast, the spin remains in the off-resonant state -- it has no time to relax before the next optical pulse is applied. By measuring the signal strength for varying $\tau_{\text{off}}$ (Fig.\ \ref{fig:spinPump}(f)), we determine an electron-spin relaxation time of  $T_1=48\pm5\ \mu$s. Our result shows that the design of the tunnel-barrier between QDs and back gate is well suited for spin-experiments on single QDs. This $T_1$ value is significantly larger compared to the GaAs QDs without the $n$-$i$-$p$-diode structure\cite{Beguin2018}. The point is that the $T_1$ time is potentially longer than the coherence time $T_2$, such that the relaxation process governing $T_1$ is unlikely to limit the coherence time $T_2$ \cite{Stockill2016}.

\section*{Discussion}
In summary, we have developed charge-tunable GaAs QDs with ultra-low charge noise. We show notable improvements of the GaAs QDs properties: optical linewidths are close-to lifetime-limited, blinking is eliminated, and long electron-spin lifetimes are achieved. From a materials perspective, the crucial advance is the new diode structure hosting GaAs QDs -- a key feature is that all the doping is incorporated in layers of low Al-concentration. In this way, the occupation of DX-centres is avoided and the AlGaAs layers are conducting at low temperatures. The concepts developed in this work can be transferred to thinner diode-structures that allow integration into photonic-crystals and other nanophotonic devices \cite{Liu2018,Kirsanske2017}. From a quantum photonics perspective, our results pave the way to bright sources of low-noise single photons close to the red part of the visible spectrum. This will facilitate the developments of both short-range networks and a hybrid QD-rubidium quantum memory. On account of the low-strain environment in GaAs QDs, our work can also open the door to prolonged electron spin coherence.

\section*{Acknowledgments}
\label{sec:acknowledgement}
The authors thank Jan-Philipp Jahn and Armando Rastelli for stimulating discussions. LZ received funding from the European Union's Horizon 2020 Research and Innovation programme under the Marie Sk\l{}odowska-Curie grant agreement No.\ 721394 (4PHOTON). MCL, CS, and RJW acknowledge financial support from NCCR QSIT and from SNF Project No.\ 200020\_156637. JR, AL, and ADW gratefully acknowledge financial support from the grants DFH/UFA CDFA05-06, DFG TRR160, DFG project 383065199, and BMBF Q.Link.X 16KIS0867. AJ acknowledge support from the European Union’s Horizon 2020 research and innovation programme under the Marie Sk\l{}odowska-Curie grant agreement No.\ 840453 (HiFig).

\section*{Author Contributions}
\label{sec:contrib}
LZ, MCL, GNN, AJ, and CS carried out the experiments. LZ, MCL, JR, and AL designed the sample. JR, LZ, ADW, and AL grew the sample. CS, LZ, GNN, and MCL fabricated the sample. LZ, MCL, CS, GNN, AJ, and RJW analyzed the data. MCL, LZ, and RJW wrote the manuscript with input from all the authors.

\section*{Competing Interests}
The authors declare no competing interests.

\section*{Data Availability}
The data that supports this work is available from the corresponding author upon reasonable request.

\section*{Code Availability}
The code that has been used for this work is available from the corresponding author upon reasonable request.

\section*{Methods}
\label{sec:methods}
\textbf{Sample fabrication:} The sample heterostructure and the quantum dots are grown by molecular beam epitaxy (MBE). The MBE setup is similar to the one described in Ref.\ \onlinecite{Ludwig2018}. The complete heterostructure of the sample is shown in Table\ \ref{table_structure}. All doped layers in AlGaAs have low Al-concentration ($<20\%$). The quantum dots are surrounded by AlGaAs with higher Al-concentration ($33\%$), to enable the growth of QDs close to rubidium-frequencies and with small fine-structure splittings \cite{Huo2013,Lobl2019}. We fabricate separate Ohmic contacts to the n$^{+}$ and p$^{++}$ layers. For the $n$-type back gate, the sample is locally etched down by $\sim 360\ $nm in a mixture of  sulfuric acid and hydrogen peroxide (concentrated H$_2$SO$_4$: 30$\%$ H$_2$O$_2$: H$_2$O $=$ 1:1:50). NiAuGe is then deposited by electron-beam evaporation (with three steps: 60 nm AuGe (mass ratio 88:12), 10 nm Ni, and 60 nm AuGe), followed by thermal annealing at 420 $^{\circ}$C for 30 s. For the $p$-type top gate, a thin contact pad consisting of Ti (3 nm)/Au (7 nm) is evaporated locally on the top surface of the sample. Both contacts are electrically connected with silver paint.

\textbf{Experimental setups:} The sample is cooled down to 4.2 K in a liquid helium cryostat. We perform photoluminescence with a $632.8$ nm He-Ne laser. The photoluminescence is collected by an aspheric objective lens (numerical aperture NA $=$ 0.71) and sent to a spectrometer. Resonance fluorescence is performed with a narrow-band laser (1 MHz linewidth), using a cross-polarisation confocal dark-field microscope\cite{Kuhlmann2013a,Jahn2015} to distinguish QD-signal from the scattered laser light. It is detected using superconducting-nanowire single-photon detectors and a counting hardware with a total timing jitter of $\sim35$ ps (full width at half maximum).

\textbf{Statistics of QD linewidths:} In our device, GaAs QDs with a small height (emission wavelength below $\sim785\ $nm) tend to have excellent optical properties. We find that more than every second QD has a close to lifetime-limited linewidth (see Supplementary Figure\ 4(a,b)). This includes QDs close to the $^{87}$Rb D$_{2}$ line ($\sim780\ $nm). For QDs larger in size (emission wavelength above $\sim785\ $nm), the QD linewidths are usually broader. The reason is probably the following: the GaAs QDs in our sample are grown by infilling nano-holes droplet-etched into a $10\ $nm-thin layer of Al$_{0.33}$Ga$_{0.67}$As (see Table\ \ref{table_structure}). The depths of the nano-holes, and therefore the heights of the QDs, typically range from $5\ $nm to $10\ $nm \cite{Huo2013,Lobl2019}. A QD emitting at higher wavelength tends to have a larger height\cite{Lobl2019}. When the height of a QD comes close to $10\ $nm, the optical properties could be affected by the Al$_{0.33}$Ga$_{0.67}$As/Al$_{0.15}$Ga$_{0.85}$As interface. A simple solution is to make the Al$_{0.33}$Ga$_{0.67}$As-layer $5\ $nm thicker. In this case, we expect good optical properties also for QDs of higher wavelengths.

\begin{table*}[b]
\begin{ruledtabular}
\caption{\label{table_structure}Sample design with relevant growth parameters.}
\begin{tabular}{lcccc} 
Material & Thickness (nm) & Temperature ($^\circ\text{C}$) & Duration (s) & Comments\\\cline{1-5} 
GaAs:C & 5 & 540 & 25.1 & p$^{++}$-doped epitaxial gate\\
Al$_{0.15}$Ga$_{0.85}$As:C & 10 & 540 & 42.7 & p$^{++}$-doped epitaxial gate\\
Al$_{0.15}$Ga$_{0.85}$As:C & 65 & 540 & 277.7 & p$^{+}$-doped epitaxial gate\\
Al$_{0.33}$Ga$_{0.67}$As & 273.6 & 540 & 921.8 & blocking barrier\\
GaAs & 2 & 605 & 10 & filling of the etched nano-holes\\
-- & -- & 605 & 60 & droplet etching\\
Al & -- & 605 & 3.7 & Al-droplet 0.9 nm plus 1 ML Al\footnotemark\\
Al$_{0.33}$Ga$_{0.67}$As & 10 & 590 & 33.7 & tunnel barrier (high Al)\\
Al$_{0.15}$Ga$_{0.85}$As & 15 & 590 & 64.1 & tunnel barrier (low Al)\\
Al$_{0.15}$Ga$_{0.85}$As & 5 & 575 & 21.4 & tunnel barrier (low-temperature)\\
Al$_{0.15}$Ga$_{0.85}$As:Si & 150 & 590 & 640.8 &  n$^{+}$-doped back gate\footnotemark\\
Al$_{0.15}$Ga$_{0.85}$As & 50 & 590 & 209.3 & buffer layer\\
AlAs/Al$_{0.33}$Ga$_{0.67}$As & 10$\times$(67.08/59.54) & 590 & 8904.7 & distributed Bragg reflector\\
GalAs/AlAs & 22$\times$(2.8/2.8) & 590 & 1101.7 & short-period superlattice\\
GaAs & 100 & 590 & 601.8 & start\\
\end{tabular}
\footnotetext[1]{For the Al-layer, the amount of deposited aluminum is given as the thickness of a corresponding AlAs-layer. The aluminum is deposited in an arsenic-depleted ambience.}
\footnotetext[2]{In the molecular beam epitaxy chamber used here, the background impurity concentration is estimated to be $\sim 5\times10^{14}\ \text{cm}^{-3}$ for Al$_{0.33}$Ga$_{0.67}$As layers\cite{MacLeod2015}. The doping concentration is $\sim 2\times10^{18}\ \text{cm}^{-3}$ for the n$^{+}$ layer, while for p$^{+}$ and p$^{++}$ layers, it is around $2\times10^{18}\ \text{cm}^{-3}$ and $8\times10^{18}\ \text{cm}^{-3}$, respectively. Between the $n$-type back gate and the $p$-type top gate, the sample has a built-in potential of $1.82\ $V.}
\end{ruledtabular}
\end{table*}

\textbf{Auto-correlation under different excitation schemes:} We investigate the stability of the QD under different excitation schemes. We start with continuous-wave (CW) excitation. We perform auto-correlation measurements on X$^{1-}$ at a constant gate voltage while exciting the QD with (i) an above-band laser ($\lambda=632.8\ $nm), (ii) a laser resonant with the $p$-shell, and (iii) a laser resonant with the $s$-to-$s$ transition. The results are shown in (i) Supplementary Figure\ 6(a), (ii) Supplementary Figure\ 6(b), and (iii) Fig.\ \ref{fig:Linewidth}(g), respectively. In all three cases, the $g^{(2)}$ stays very flat and close to one -- there is no blinking even on a long time-scale. This shows that the QD is a very stable quantum emitter under all three CW excitation schemes. From an applications point of view, it is usually necessary to drive the QD with a resonant pulsed laser. We investigate the auto-correlation under resonant $\uppi$-pulse excitation in Fig.\ \ref{fig:Linewidth}(e). An evaluation of this $g^{(2)}$-measurement on a longer time-scale is plotted in Supplementary Figure\ 6(c), where the $y$-axis is displayed on a logarithmic scale to resolve the central peak. To investigate whether a strong $\uppi$-pulse introduces any blinking, we plot the $g^{(2)}$-measurement in a histogram plot (Supplementary Figure\ 6(d)) by summing up the coincidence events for every single pulse. This sum is divided by the expectation value for a perfectly stable source: the normalisation factor is $x_1x_2T_{\text{int}}/f_{\text{rep}}$, where $f_{\text{rep}}$ is the repetition rate of the pulsed laser, $x_1$, $x_2$ represent the count rates of the two detectors used for a $T_{\text{int}}$-long $g^{(2)}$-measurement. A derivation of the normalisation factor is given in Supplementary Figure\ 6. Importantly, the histogram bars at non-zero time delay are flat and very close to one; the bar at zero delay is close to zero. This shows that the QD is a stable single-photon emitter for resonant $\uppi$-pulse excitation.

\textbf{Potential noise source affecting the QD-linewidth:} The $g^{(2)}$-measurement shown in Fig.\ \ref{fig:Linewidth}(f,g) is performed on a trion at zero magnetic field when the CW laser drives the QD resonantly. The sensitivity can be enhanced towards either spin noise or charge noise by applying a small magnetic field along the growth direction, $B$, and detuning the laser slightly from the QD-resonance by $\delta$, respectively. A trion state is degenerate at zero magnetic field, consisting of two opposite spin ground states. When applying a magnetic field $B$, the degeneracy is lifted and the trion state is split into two by a Zeeman energy $E_z=g\mu_BB$, with $g$ being the electron or hole $g$-factor, and $\mu_B$ the Bohr magneton. We maximise the spin noise sensitivity by applying a small magnetic field such that $E_z=\frac{\tilde{\Gamma}}{\sqrt{3}}$ (Supplementary Figure\ 8(c)). Here $\tilde{\Gamma}$ represents the full width at half maximum (FWHM) of the QD emission. For the maximised spin noise sensitivity, the $g^{(2)}$-measurement does not show any clear sign of bunching (Supplementary Figure\ 8(d)). The charge noise sensitivity is maximised when the laser is detuned from the QD by $\delta=\frac{\tilde{\Gamma}}{2\sqrt{3}}$ (Supplementary Figure\ 8(e)). In this configuration, we observe a small bunching peak in the $g^{(2)}$-measurement (Supplementary Figure\ 8(f)). This result suggests that charge noise on a millisecond time-scale is responsible for the slight linewidth broadening. 

\bibliography{lit}

\onecolumngrid
\clearpage
\section*{Supplementary Information}
\label{sec:Supplement}
\renewcommand{\figurename}{Supplementary Figure}
\setcounter{figure}{0}
\twocolumngrid
\begin{figure*}
\includegraphics[width=0.6\columnwidth]{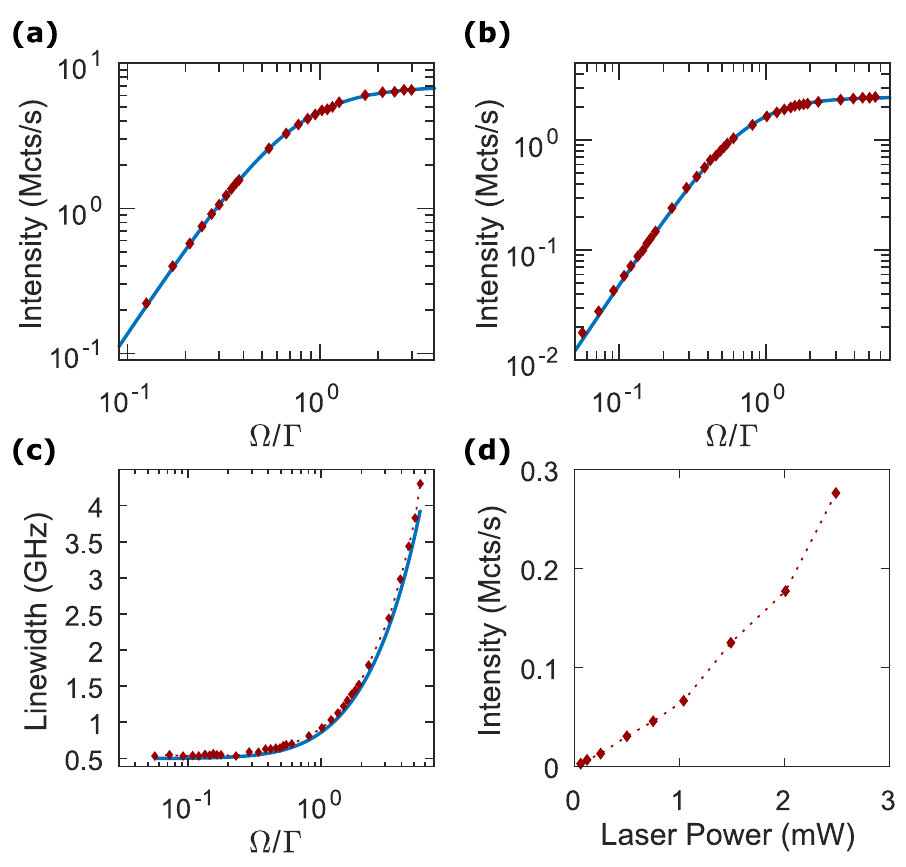}
\caption{\label{fig:powerseries}{\bf Power dependent measurements on QD1 and QD2.}{\bf (a)} Intensity of the resonance fluorescence (red diamonds) of X$^{1-}$ from QD1 as a function of the normalised Rabi-frequency $\Omega/\Gamma$. The resonance fluorescence is measured using a dark-field confocal microscope\cite{Kuhlmann2013a}. The intensity saturates at around $6.5$ Mcts/s (raw count-rate). The blue curve is a theoretical fit of a two-level model \cite{Loudon2000}. {\bf (b)} Power dependent resonance fluorescence measurement (red diamonds) of X$^{1-}$ from QD2. The measurement is performed similar to the one shown in (a) and recorded with a superconducting nanowire single-photon detector. The intensity saturates around $2.5$ Mcts/s (raw count-rate). The blue curve represents again a theoretical fit of a two-level model. {\bf(c)} Optical linewidth (red) of the resonant fluorescence from QD2 X$^{1-}$ displayed as a function of the normalised Rabi-frequency $\Omega/\Gamma$. The linewidth measurements are performed by scanning the gate voltage acorss the QD resonance under different excitation laser powers (the laser frequency is fixed). The linewidths are fitted to Lorentzian functions and converted into frequency unit using a Stark shift of $621.679$ GHz/V (see Fig.\ 2(c) of the main text). The linewidths stay very close to the lifetime limit ($496$ MHz) at low power and become broader due to power broadening at higher power. The blue curve represents the power broadening effect of a two-level system. The plot is not a fit but just the theoretical model using the parameters $\Gamma$ as well as $\Omega$ extracted from (b). The measured optical linewidths stay close (marginally above) to theoretical values for all excitation powers. {\bf (d)} Intensity of photoluminescence of X$^{1-}$ from the same QD, plotted as a function of non-resonant laser power. This non-resonant laser is a CW He-Ne laser emitting at $\lambda=632.8$ nm. The non-resonant laser power is measured before we send it to the confocal microscope. Photoluminescence is measured at the same gate voltage as in (b). It is collected by the microscope setup, sent through a grating-based filter ($50$ GHz bandwidth, for filtering out non-resonant laser) and counted by the superconducting nanowire single-photon detector. Under lower-power non-resonant excitation, we determine an upper bound of $7.7$ GHz for the photoluminescence linewidth. This boundary is obtained by fitting the photoluminescence spectrum to a Lorentzian function, and is limited by the resolution of the spectrometer. The actual photoluminescence linewidth could be much narrower.}
\end{figure*}

\begin{figure*}
\includegraphics[width=1.0\columnwidth]{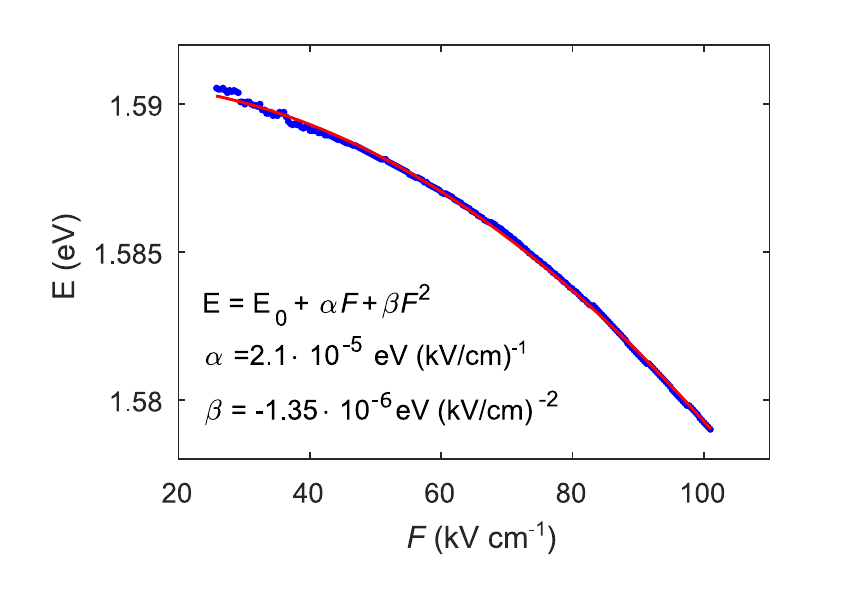}
\caption{{\bf Photoluminescence as a function of electric field.} Photoluminescence energy ($E$) versus electric field\cite{Warburton2002} ($F$) for the quantum dot shown in Fig.\ 1 of the main text. The electric field is obtained by a bandstructure simulation. The solid red curve is a quadratic fit to the data. We extract the permanent dipole moment to be $\alpha/e = 0.21$ nm, and the polarisability $\beta = -1.35\ \mu$eV\ (kV/cm)$^{-2}$.}.
\end{figure*}

\begin{figure*}
\centering
\includegraphics[width=2 \columnwidth]{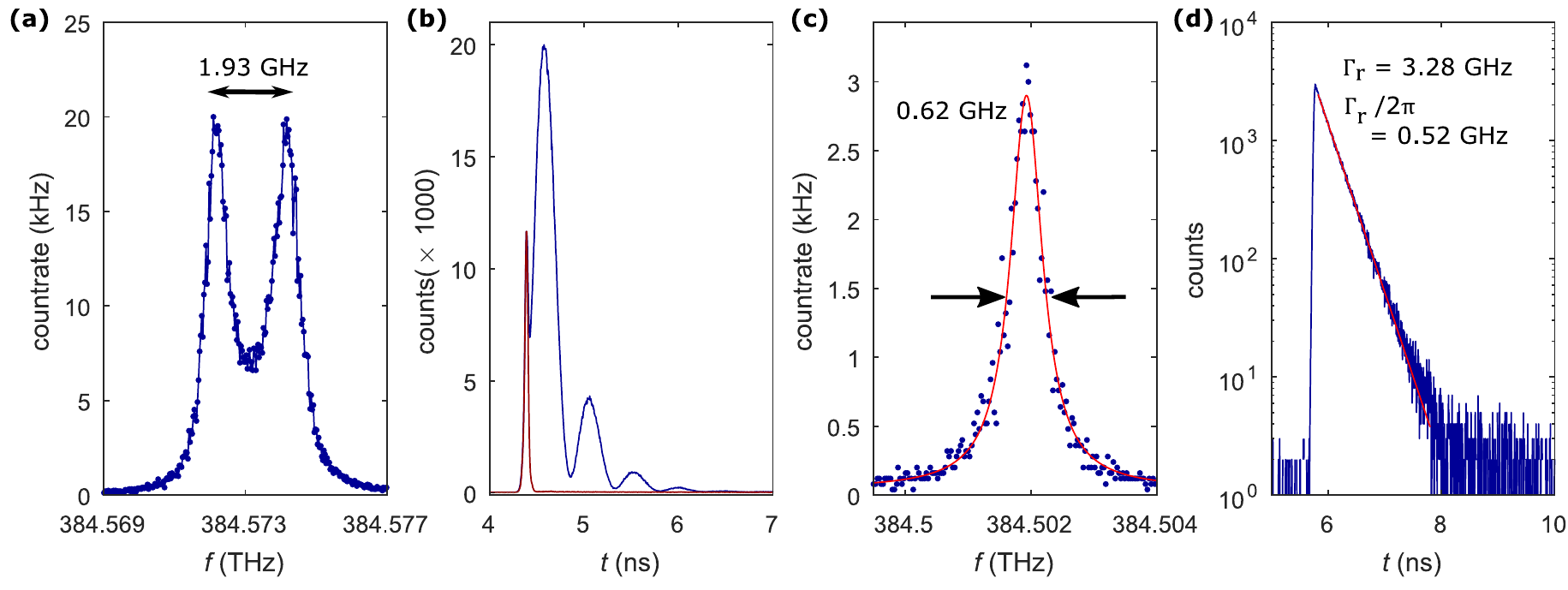}
\caption{\label{fig:X0}{\bf Resonant linewidth and lifetime measurements on X$^0$ and X$^+$.}{\bf (a)} Resonance fluorescence from the neutral exciton, X$^0$, measured on QD2. The neutral exciton has a small fine structure splitting of FSS = $1.93$ GHz ($7.98\ \mu$eV). {\bf (b)} Lifetime measurement on the X$^0$ (QD2). Resonance fluorescence (blue) is measured as a function of the time delay $t$ after exciting X$^0$ with a picosecond laser-pulse. Here, a pronounced quantum beat with a frequency of $\frac{2\uppi}{\text{FSS}}$ is observed. The red curve corresponds to the background from the scattered laser light. {\bf (c)} Resonance fluorescence measurement (blue) on the positively charged exciton, X$^+$, from QD2. The red curve is a Lorentzian fit. The X$^+$ shows a narrow optical linewidth of $0.62$ GHz. {\bf (b)} Lifetime measurement (blue) on the X$^+$ (QD2). The radiative decay rate, which corresponds to a natural linewidth of $\Gamma_{\text{r}}/2\uppi=0.52$ GHz, is extracted by fitting an exponential curve (red).}
\end{figure*}

\begin{figure*}[b]
\includegraphics[width=1\columnwidth]{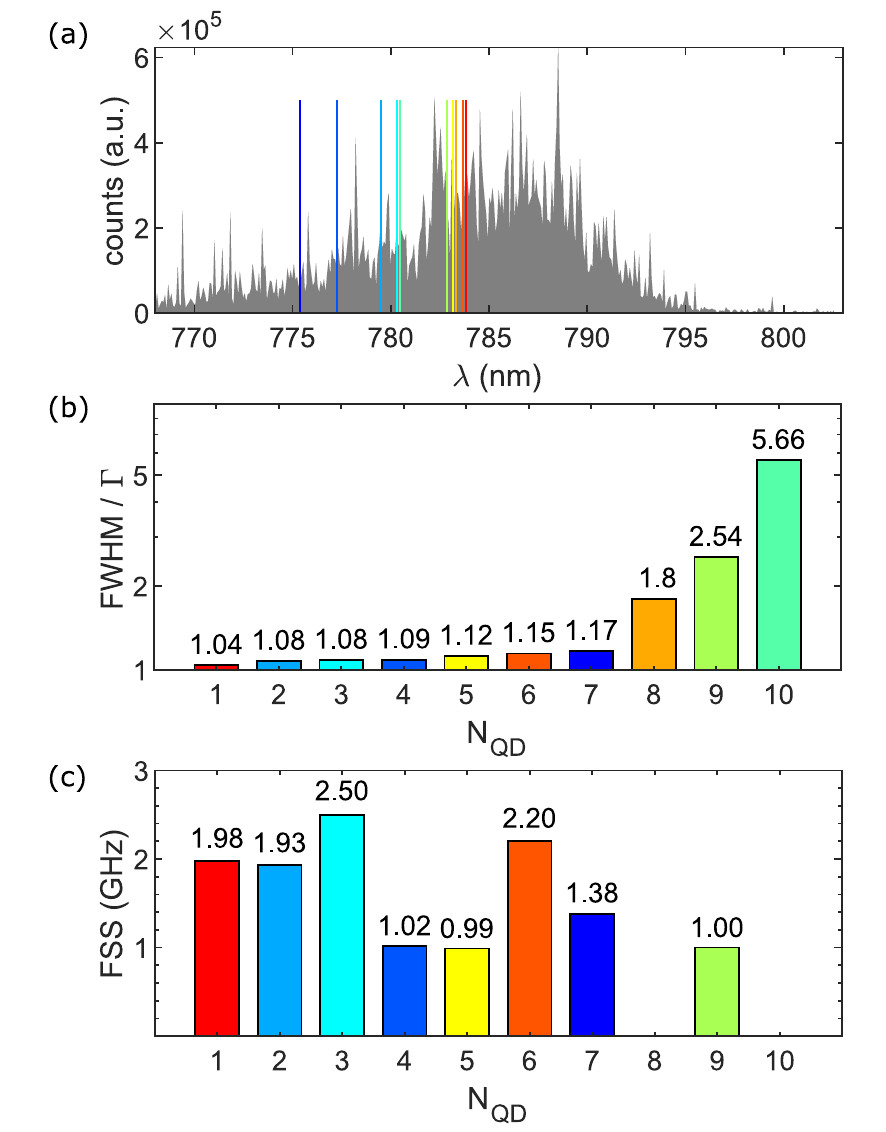}
\caption{\label{fig:lw_FSS}{\bf Summary of the optical properties of ten randomly chosen QDs. (a)} Emission of the QD-ensemble at a gate voltage of $V_{\text{g}}=-0.4\ $V. The coloured lines indicate the emission wavelengths of the QDs (X$^{1-}$) which have been measured in detail. {\bf(b)} The ratio between measured optical linewidth (full width at half maximum, FWHM) on the negatively charged exciton X$^{1-}$ and its lifetime limit ($\Gamma$) shown for ten randomly chosen QDs. On the $x$-axis, N$_{\text{QD}}$ indicates the QD number sorted by the ratio FWHM$/\Gamma$ in the ascending order. The colours of the bars are linked to the colours in (a). In the ideal case, the optical linewidth of QD reaches the lifetime limit: the ratio FWHM$/\Gamma$ is one. For the majority of QDs, the ratio is close to one -- below a level of FWHM$/\Gamma$ = 1.2 we find seven QDs out of ten. These QDs suffer from little noise. The QD1 and QD2 investigated in the paper are labelled here as N$_{\text{QD}}$ = 3 and N$_{\text{QD}}$ = 2, respectively. For few QDs, there is a rather large broadening of the linewidth beyond the lifetime limit. The lifetime limits of QD1, QD4 - QD10 are 510 MHz, 640 MHz, 515 MHz, 520 MHz, 437 MHz, 496 MHz, 250 MHz, 530 MHz, respectively. {\bf(c)} Fine structure splitting (FSS) for the neutral exciton (X$^{0}$) measured on the same QDs as in (b). The FSS is determined by scanning the laser frequency across the QD resonance. An example is shown in Supplementary Figure\ 3. As in (b), the colour of the bars is linked to the QD wavelengths in (a). For most of the QDs, the FSS is below 2 GHz. For N$_{\text{QD}}$ = 9, we state here an upper bound of 1 GHz for its FSS. For N$_{\text{QD}}$ = 8$\ \text{\&}\ $10, determining the FSS by scanning a laser across X$^{0}$ was not successful due to the relatively large linewidth.}
\end{figure*}

\begin{figure*}
\includegraphics[width=2.0\columnwidth]{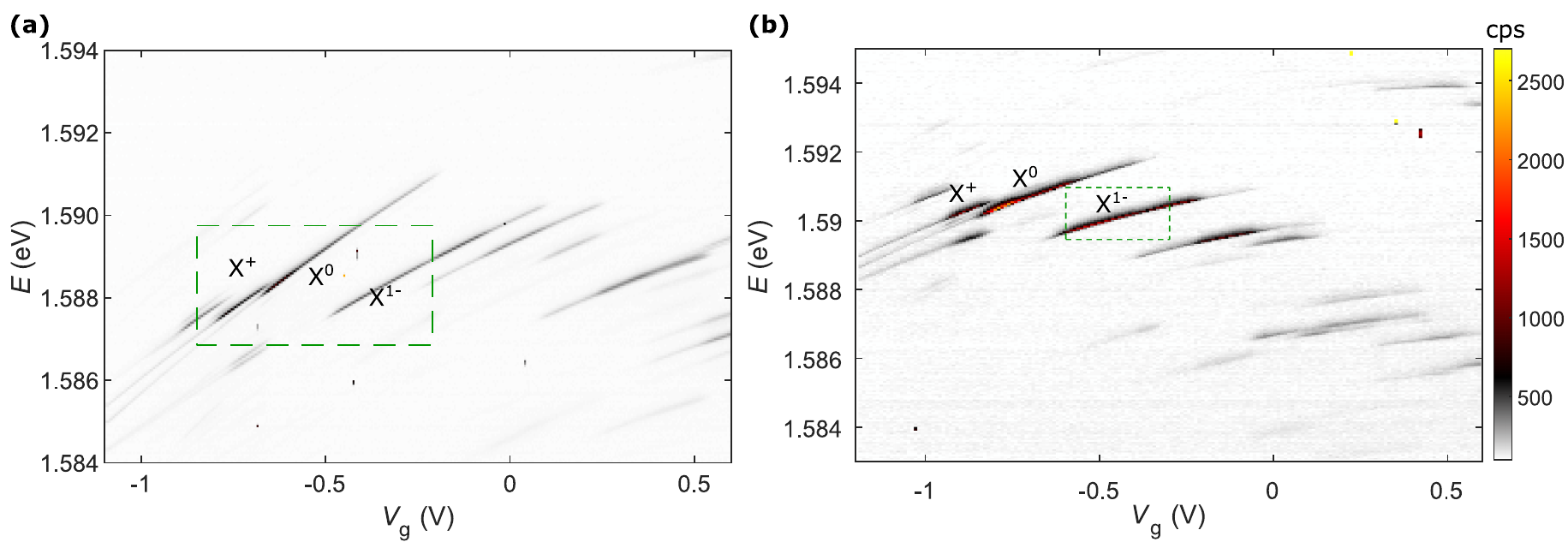}
\caption{\label{fig:RFplateau} {\bf Photoluminescence charge plateaus of QD1 and QD2.} {\bf (a)} Charge plateaus of QD1 measured in photoluminescence. We observe emission over a wide range of excitons with different net-charges. The green dashed frame indicates the scan-range of the measurement in Fig.\ 1(c) of the main text. {\bf(b)} Similar photoluminescence measurement on QD2. Again, several charge plateaus are observed. The green dashed area indicates the scan-range of the resonance fluorescence measurement in Fig.\ 2(c) of the main text.}
\end{figure*}

\begin{figure*}[b]
\includegraphics[width=1.8\columnwidth]{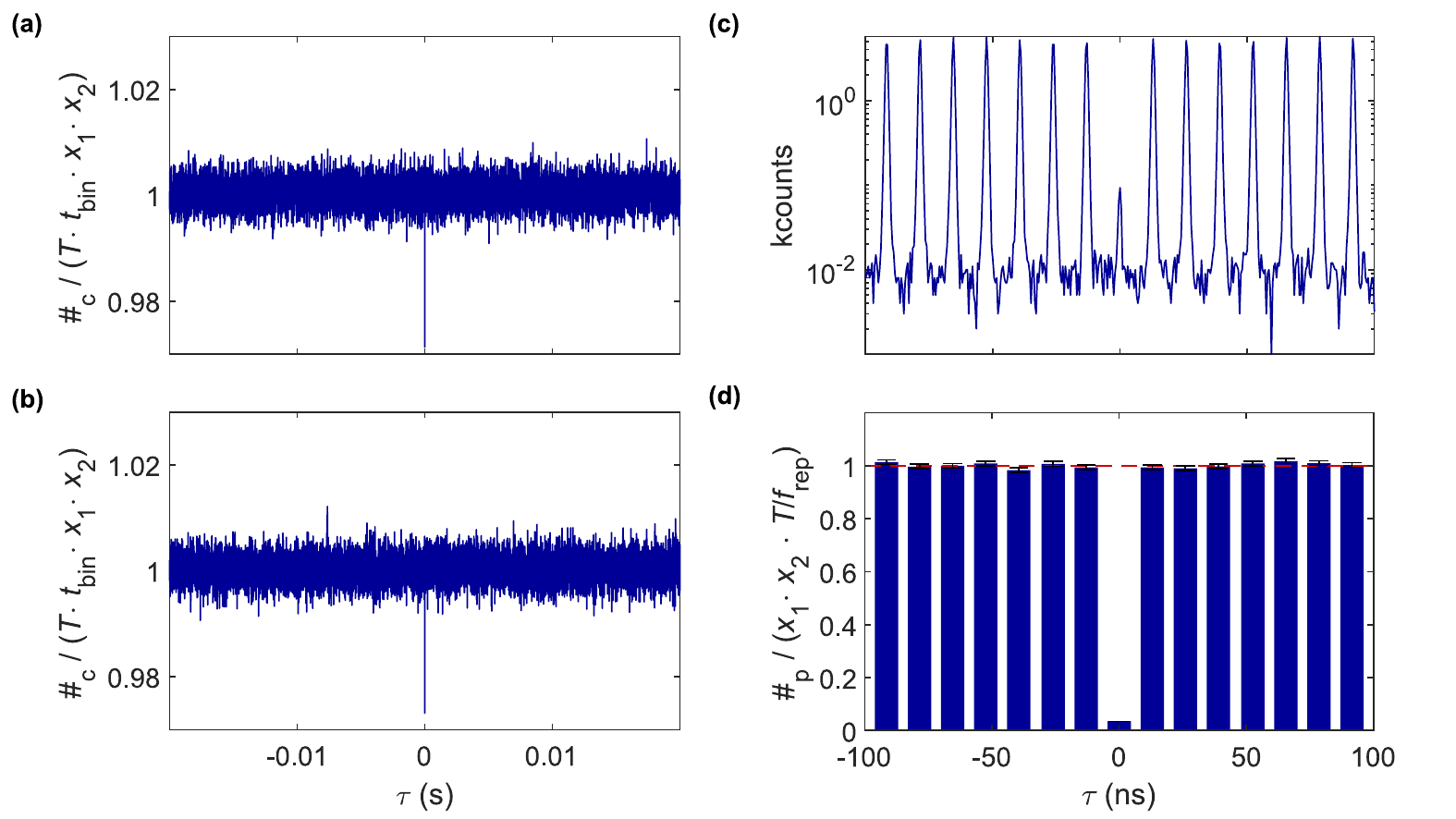}
\caption{\label{fig:Ex2}  {\bf Auto-correlation of the QD1 emission (X$^{1-}$) from $s$-to-$s$ recombination for non-resonant and pulsed excitation. (a)} The auto-correlation measurement performed under above-band excitation with a laser at $\lambda=632.8\ $nm. The $g^{(2)}(\tau)$ is close to the ideal Poissonian limit of one for all time-scales, demonstrating the long-time stability of the emitter. {\bf (b)} The auto-correlation measurement performed by exciting the quantum dot via $p$-shell excitation. Similarly here, the $g^{(2)}$ is flat and close to the ideal Poissonian limit. {\bf (c)} The pulsed $g^{(2)}$-measurement (resonant $\uppi$-pulse excitation) from Fig. 2(e) plotted on a logarithmic scale and evaluated on a longer time-scale. The offset (about ten coincidence events) arises from detector dark counts. For the calculation of the $g^{(2)}(0)$ value in the main text ($g^{(2)}(0)=0.019$), we have subtracted this dark-counts induced background. By integrating all coincidence events over one full pulse period without substrating the background, we estimate a ``worst-case" upper limit: $g^{(2)}(0)=0.036$. {\bf (d)} The same $g^{(2)}$-measurement as in (c) but plotted as a histogram. To obtain the histogram, we sum up all coincidence events within every single pulse and normalise it by the expectation value for the coincidence events in the case of an ideal Poissonian source: $\langle\#_{\text{p}}\rangle=x_1x_2T_{\text{int}}/f_{\text{rep}}$, where $x_1$ and $x_2$ are the count rates of each detector channel, $T_{\text{int}}=2500\ $s is the overall integration time for the measurement, and $f_{\text{rep}}=76.36\ $MHz is the repetition rate of the pulsed laser. Using this normalisation factor, the $g^{(2)}$-measurement normalises to a value very close to the ideal limit of one. This evaluation shows that the quantum dot is a very stable emitter under resonant $\uppi$-pulse excitation. The normalisation factor $\langle\#_{\text{p}}\rangle$ is obtained by a similar consideration compared to the case of continuous-wave excitation (see Ref. \onlinecite{Lobl2020}): let $p_1$ and $p_2$ be the probabilities that a photon is detected on channel 1 or 2 after the $\uppi$-pulse excitation. The count-rates $x_1$, $x_2$ are then connected to these probabilities by $x_1=f_{\text{rep}}p_1$, $x_2=f_{\text{rep}}p_2$. In case of two uncorrelated channels, the joint probability for one detection event on channel 1 together with another detection event for a later or earlier excitation pulse on channel 2 is $p_1p_2=x_1x_2/f_{rep}^2$. The expectation value for the overall number of joint (coincidence) events is then $p_1p_2$ times the overall number of $\uppi$-pulses, $T_{\text{int}}\cdot f_{\text{rep}}$. This consideration leads to the aforementioned expression: $\langle\#_{\text{p}}\rangle=x_1x_2/f_{\text{rep}}^2\times T_{\text{int}}\cdot f_{\text{rep}} = x_1x_2T_{\text{int}}/f_{\text{rep}}$.}
\end{figure*}

\begin{figure*}
\centering
\includegraphics[width=2.0\columnwidth]{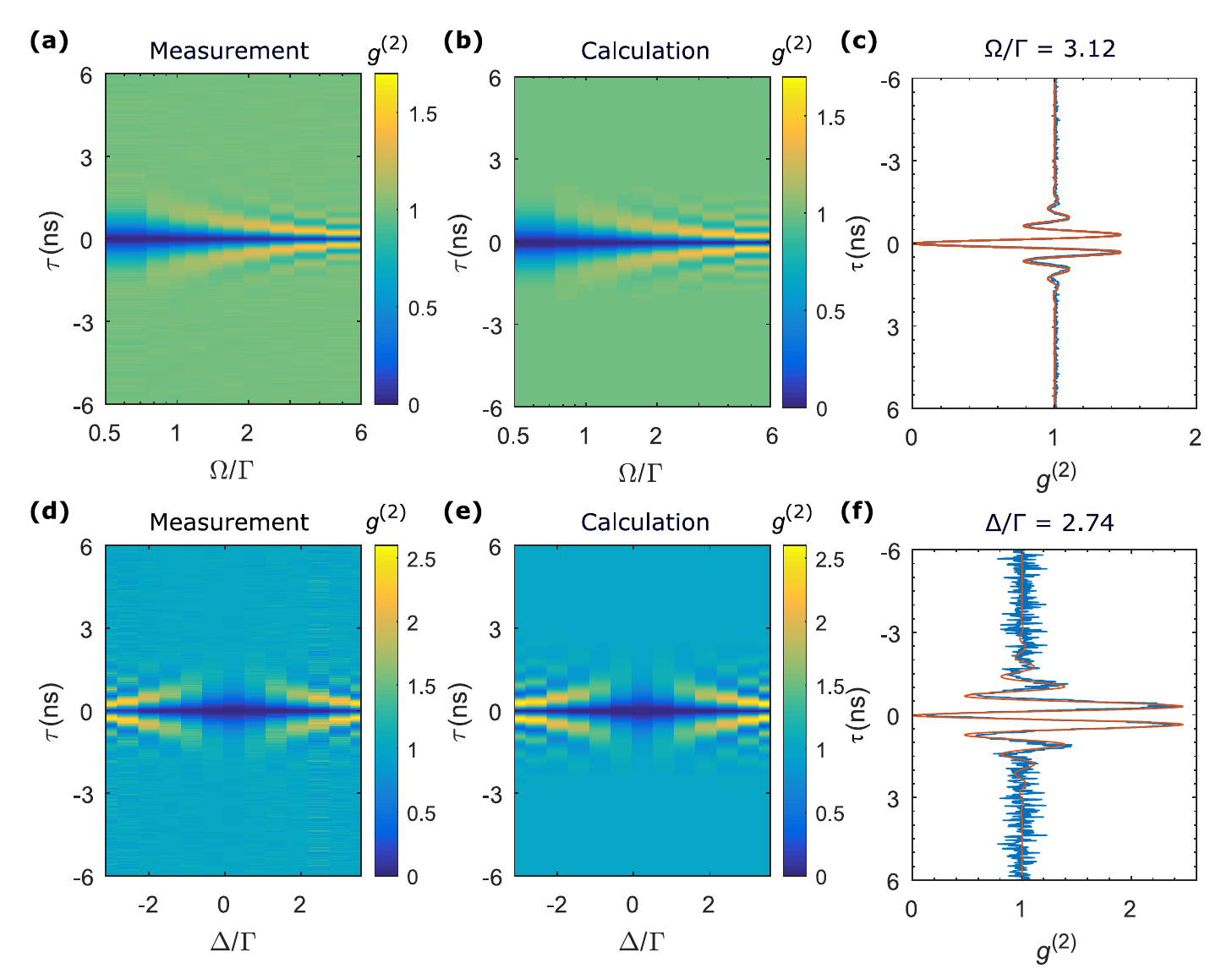}
\caption{\label{fig:g2Detune}{\bf Power-dependent and detuning-dependent autocorrelation measurements.} {\bf (a)} Time-resolved intensity autocorrelation measurement, $g^{(2)}(\tau)$, as a function of the normalised Rabi-frequency $\Omega/\Gamma$. The measurement is performed under continuous wave (CW) resonant excitation on the negatively charged exciton, X$^{1-}$, from QD2. The $g^{(2)}(\tau)$ is normalised to one by the number of coincidence events at 300 ns delays. The Rabi frequency $\Omega$ is extracted independently from a power saturation curve under CW resonant excitation, while the radiative decay rate $\Gamma$ is obtained by fitting an exponential function to the lifetime measurement. {\bf (b)} Calculation of the power-dependent autocorrelation function. The $g^{(2)}(\tau)$ function is calculated by solving the optical Bloch equations of a two-level system and then applying the quantum regression theorem. The calculation is carried out with QuTip\cite{Johansson2013}. We chose to ignore upper-level dephasing in this calculation\cite{Jahn2015}. The normalised Rabi-frequency is taken from the measurement in {\bf (a)}. Under a strong driving field ($\Omega \gg \Gamma$), the $g^{(2)}(\tau)$ value approaches its upper bound\cite{Flagg2009} $g^{(2)}(\tau = \pm\uppi/\Omega) = 2$.  {\bf (c)} Comparison between the measured $g^{(2)}(\tau)$ (blue) and the calculation (red) for $\Omega/\Gamma = 3.12$. {\bf (d)} Time-resolved intensity autocorrelation measured as a function of normalised laser-detuning, $\Delta/\Gamma$. The excitation laser power is locked to $\Omega = 0.49\ \Gamma$. {\bf (e)} Calculation of the detuning-dependent autocorrelation function in a two-level system. Under a detuned driving, the effective Rabi-frequency\cite{Rezai2019,Loudon2000} is represented as $\sqrt{\Omega^2+\Delta^2}$. In the calculation, the dephasing is again set to zero, and the values of $\Delta/\Gamma$ are taken from (d). In both the measurement and the calculation, the maximum value of $g^{(2)}(\tau)$ exceeds 2, the upper bound in the resonant case, when $\Delta$ is relatively large compared to $\Gamma$. {\bf (f)} Comparison between the experiment (blue) and the calculation (red) under the condition $\Delta/\Gamma = 2.74$, $\Omega/\Gamma = 0.49$. We find a very good overlap between the data and the calculation curve, indicating the X$^{1-}$ behaves here as an ideal two-level system.}
\end{figure*}

\begin{figure*}[b]
\includegraphics[width=1.8\columnwidth]{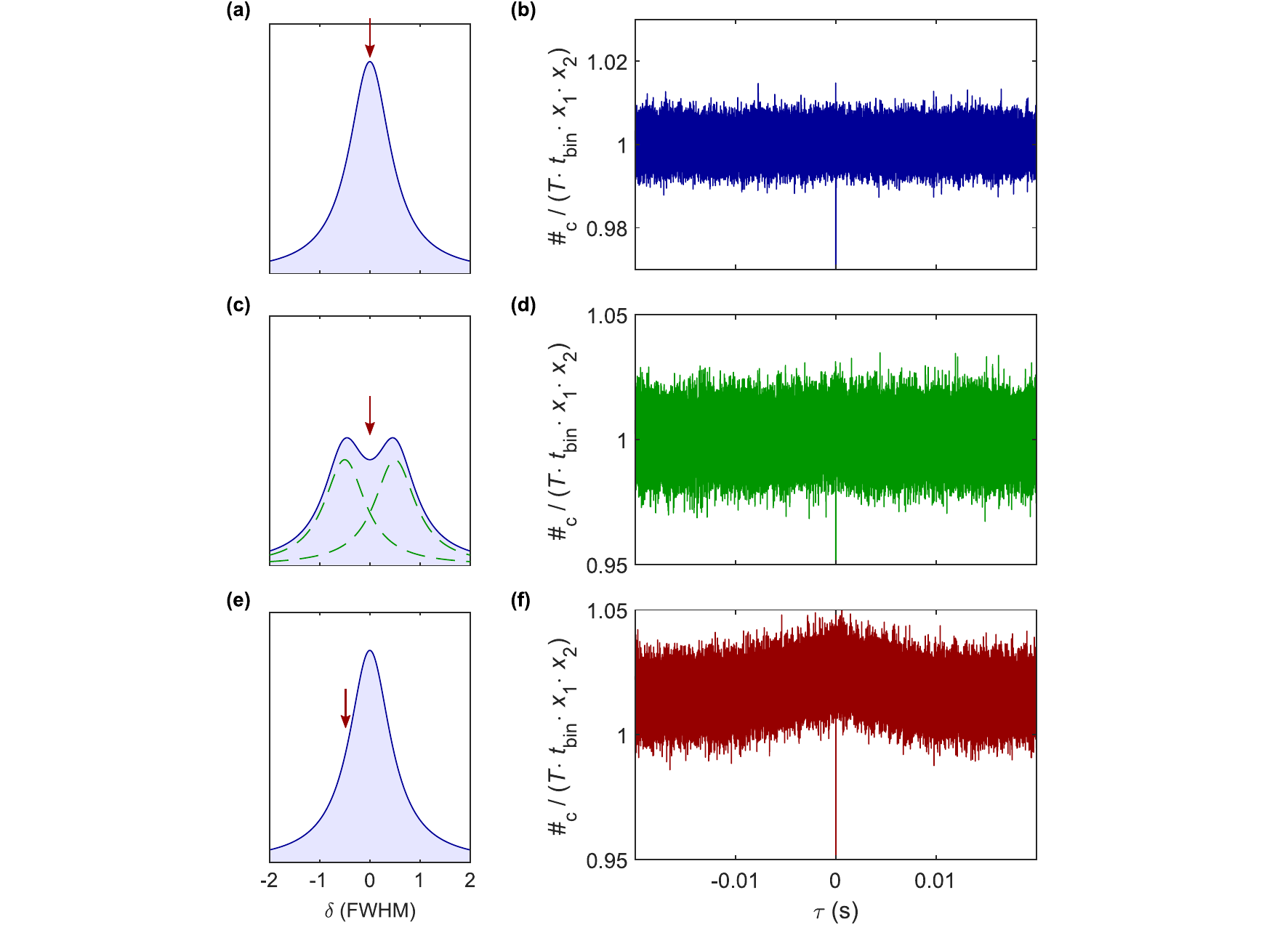}
\caption{\label{fig:Ex1}{\bf Auto-correlation of the resonance fluorescence on QD1 X$^{1-}$.} {\bf (a)} Configuration for resonant auto-correlation measurement. The measurement is performed at zero magnetic field ($B=0\ $T) with the laser on resonance with the quantum dot ($\delta=0$). {\bf (b)} Result of the auto-correlation measurement, $g^{(2)}(\tau)$, as described in (a) evaluated for long time-scales. The $g^{(2)}(\tau)$ is perfectly flat and stays close to one. {\bf (c)} Configuration for an auto-correlation measurement with enhanced sensitivity to spin noise. This measurement is performed at a finite magnetic field ($B=20\ $mT) along the growth direction with the laser frequency centred between the two Zeeman peaks. {\bf (d)} Result of the $g^{(2)}$-measurement as described in (c) for long time-scales. The $g^{(2)}(\tau)$ remains flat and close to one. {\bf (e)} Configuration for an auto-correlation measurement with enhanced sensitivity to charge noise. This measurement is performed at zero magnetic field with the laser slightly detuned (by about half of the linewidth) with respect to the quantum dot resonance. {\bf (f)} Result of the $g^{(2)}$-measurement as described in (e). Here, a bunching on a millisecond time-scale can be seen, showing that some charge noise is present. The charge noise is likely to account for the residual linewidth broadening of X$^{1-}$.}
\end{figure*}

\begin{figure*}
\includegraphics[width=2.0\columnwidth]{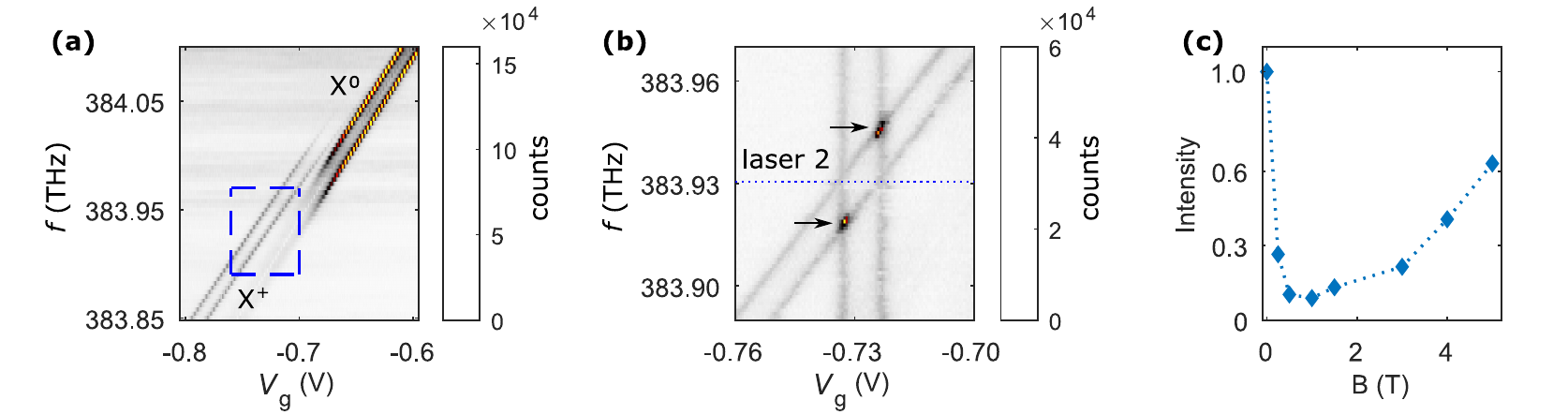}
\caption{\label{fig:Xplus} {\bf Spin pumping on the positively charged trion.}{\bf (a)} Resonance fluorescence charge-plateau of the positively charged trion, X$^{1+}$, and the edge of the X$^0$ charge-plateau (QD1). The X$^{1+}$ lines are split in a magnetic field ($B = 1.5$ T). The resonance fluorescence is weak due to optical spin-initialisation of the hole spin. {\bf (b)} The signal recovers (marked with arrows) on addressing a second spin ground state with a second laser (dashed line). This second laser is kept at a fixed frequency and a fixed power (same power as in (a)). The blue frame in (a) indicates the range over which the gate voltage and the first laser is tuned in this measurement. Two additional vertical lines are observed when the fixed laser is on resonance with the two vertical transitions. {\bf (c)} The brightness of the X$^{1+}$ resonance fluorescence as a function of the magnetic field. The brightness is normalised to the resonance fluorescence intensity at $B = 0\ \text{T}$. At about 1 T, the signal has a minimum, suggesting that the lifetime of the hole-spin is the longest at this magnetic field \cite{Dreiser2008}.}
\end{figure*}

\end{document}